\documentclass[10pt]{IEEEtran}
\usepackage{graphics}
\usepackage[dvips]{graphicx}
\newcommand{\qed}{\fbox{}}
\newcommand{\ve}[1]{ \mbox{\boldmath$#1$} }
\newcommand{\defeq}{\stackrel{\triangle}{=}}

\newtheorem{example}{Example}
\newtheorem{definition}{Definition}

\newtheorem{lemma}{Lemma}
\newtheorem{corollary}{Corollary}
\newtheorem{theorem}{Theorem}
\newtheorem{remark}{Remark}

\newcommand{\Gen}{{\cal G}}
\newcommand{\G}{{\cal G}}

\newcommand{\Ren}{{\cal R}}

\newcommand{\Ben}{{\cal B}}
\newcommand{\supp}{{\rm Supp}}

\title{On Undetected Error Probability \\ of Binary Matrix Ensembles} 

\author{Tadashi Wadayama$^\dagger$\thanks{$\dagger$Nagoya Institute of Technology, email:wadayama@nitech.ac.jp. A part of this work was presented at ITA workshop in UCSD, Feb. 2007.}}
 
\begin{document}
\maketitle

\begin{abstract}
In this paper, an analysis of the undetected error probability of 
ensembles of $m \times n$ binary matrices  is presented.
The ensemble called the {\em Bernoulli ensemble} whose 
members are considered as matrices generated from  i.i.d. Bernoulli source 
is mainly considered here.
The main contributions of this work are 
(i) derivation of the error exponent of the average undetected error probability 
and (ii) closed form expressions for the variance of the undetected error probability.
It is shown that the behavior of the exponent for a sparse ensemble is somewhat different from that for a 
dense ensemble. Furthermore, as a byproduct of the proof of the variance 
formula, simple covariance formula of the weight distribution is derived. 
\end{abstract}

\section{Introduction}
 {\em Random coding} is an extremely powerful technique 
to show the existence of a code satisfying certain properties.
It has been used for proving the direct part (achievability) of many 
types of coding theorems. 
Recently, the idea of random coding has also come to be regarded as important from a practical point of view. 
An LDPC (Low-density parity-check) code can be constructed by choosing a parity check 
matrix  from an ensemble of sparse matrices.
Thus, there is a  growing interest in randomly generated codes. 

One of the main difficulties associated with the use of randomly generated codes is the difficulty in evaluating 
 the properties or performance of such codes. For example, it is difficult to evaluate 
 minimum distance, weight distribution, 
ML decoding performance, etc. for these codes. 
To overcome this problem, we can take a {\em probabilistic approach}.
In such an approach, we consider an ensemble of parity check matrices:  
i.e., probability is assigned to each matrix in the ensemble. 
A property of a matrix (e.g., minimum distance, weight distributions) can then be regarded as
a random variable.  It is  natural to consider statistics of the random variable such as
mean, variance, higher moments and covariance. In some cases, we can show that
a property is strongly concentrated around its expectation. Such a concentration result justifies
the use of the probabilistic approach.

Recent advances in the analysis of the average weight distributions of 
LDPC codes, such as those described by Litsyn and Shevelev \cite{LS02}\cite{LS03}, Burshtein and Miller \cite{MV04},
Richardson and Urbanke \cite{modern}, show that
the probabilistic approach  is a useful technique for investigating
typical properties of codes and matrices, which are not easy to obtain.
Furthermore, the second moment analysis of the weight distribution of LDPC codes \cite{BB05}\cite{VR05} 
can be utilized to prove concentration results for weight distributions.

The evaluation of the error detection probability of a given code (or given parity check matrix) is 
a classical problem in coding theory \cite{klove2}, \cite{klove}  and some results on this topic have been derived from
the view point of a probabilistic approach.  For example, for a linear code ensemble the inequality,
$P_U<2^{-m}$ has long been known where $P_U$ is the undetected error probability 
and $m$ is the number of rows of a parity check matrix. 
Since the undetected error probability can be expressed as a linear combination of 
the weight distribution of a code, there is a natural connection between the expectation of the
weight distribution and the expectation of the undetected error probability.

In this paper, an
analysis of the undetected error probability of ensembles of  binary matrices 
of size $m \times n$ is presented.
An error detection scheme is a crucial part of a feedback error correction scheme such as 
ARQ(Automatic Repeat reQuest).
Detailed knowledge of the error detection performance of  a matrix ensemble
would be useful for assessing the performance of  a feedback error correction scheme.

\section{Average undetected error probability}

\subsection{Notation}

For a given $m \times n (m,n \ge 1)$ binary parity check matrix $H$, 
let $C(H)$ be the binary linear code of length $n$ defined by $H$, namely,
$
C(H) \defeq \{\ve x \in F_2^n: H \ve x^t = 0^m\}
$
where $F_2$ is the Galois field with two elements $\{0,1\}$ 
(the addition over $F_2$ is denoted by $\oplus$).
The notation $0^m$ denotes the zero vector of length $m$.
In this paper, a boldface letter, such as $\ve x$ for example, denotes a binary row vector.

Throughout the paper, 
a binary symmetric channel (BSC) with crossover probability $\epsilon$
($0 < \epsilon < 1/2$) is assumed.
We assume the conventional scenario for error detection: A transmitter 
sends a codeword $\ve x \in C(H)$ to a receiver via a BSC with crossover probability $\epsilon$.
The receiver obtains a received word $\ve y = \ve x \oplus \ve e$, where $\ve e$ denotes an 
error vector. The receiver firstly computes the syndrome $\ve s = H \ve y^t$ and then
checks whether $\ve s =  0^m$ holds or not.

An undetected error event occurs when $H \ve e^t  = 0^m$ and $\ve e \ne 0^m$.
This means that the error vector $\ve e \in C(\ve e \ne 0^n)$ causes an 
undetected error event.  Thus, the undetected error probability $P_U(H)$ can be 
expressed as
\begin{equation}
P_U(H) 
= \sum_{\ve e \in C(H), \ve e \ne 0^m} \epsilon^{w(\ve e)} (1-\epsilon)^{n - w(\ve e)} 
\end{equation}
where $w(\ve x)$ denotes the Hamming weight of vector $\ve x$.
The above equation can be rewritten as
\begin{equation}
P_U(H) = \sum_{w = 1}^n A_w(H) \epsilon^w (1-\epsilon)^{n - w},
\end{equation}
where $A_w(H)$ is defined by
\begin{equation}
A_w(H) \defeq \sum_{\ve x\in Z^{(n,w)}} I[H \ve x^t = 0^m].
\end{equation}
The set $\{A_w(H) \}_{w=0}^n$ is usually called the { \em weight distribution} of $C(H)$.
The notation $Z^{(n,w)}$ denotes the set  of $n$-tuples with weight $w$.
The notation $I[condition]$ is the indicator function
such that $I[condition] = 1$ if $condition$ is true; otherwise, it evaluates to 0.

Suppose that $\Gen$ is a set of binary $m\times n$ matrices $(m, n \ge 1)$.
Note that  $\Gen$ may contain some matrices with all elements identical.
Such matrices should be distinguished as distinct matrices. 
A probability $P(H)$ is associated with each matrix $H$ in $\Gen$.  Thus, 
$\Gen$ can be considered as an {\em ensemble} of binary matrices.
Let $f(H)$ be a real-valued function which depends on $H \in \Gen$.
The expectation  of $f(H)$ with respect to the ensemble $\Gen$ is defined by
\begin{equation}
E_{\Gen}[f(H)] \defeq \sum_{H \in \Gen} P(H) f(H).
\end{equation}
The average weight distribution of a given ensemble $\Gen$ is given by
$
E_{\Gen}[A_w(H)].
$
This quantity is very useful for analyzing the performance of binary linear codes, including analysis of
the undetected error probability.

\subsection{Bernoulli ensemble}

In this paper, we will focus on a parameterized ensemble $\Ben_{m,n,k}$
which is called the {\em Bernoulli ensemble}
because the Bernoulli  ensemble is amenable to ensemble analysis.
The Bernoulli  ensemble $\Ben_{m,n,k}$ contains 
all the binary $m \times n$ matrices ($m,n \ge 1$), whose elements  are 
regarded as  i.i.d. binary random variables such that 
an element  takes the value 1 with probability $p \defeq k/n$.
The parameter $k (0 < k \le n/2)$ is a positive real number which represents the 
average number of ones for each row.
In other words, a matrix $H \in \Ben_{m,n,k}$ can be considered as an output 
from the Bernoulli source such that symbol 1 occurs with probability $p$.

From the above definition, it is clear that 
a matrix $H \in \Ben_{m,n,k}$ is associated with the probability
\begin{equation}
P(H) = p^{\bar w(H)} (1-p)^{m n - \bar w(H)} ,
\end{equation}
where $\bar w(H)$ is the number of ones in $H$ (i.e., Hamming weight of $H$).
The average weight distribution of  the Bernoulli ensemble is
given by
\begin{equation}
E_{\Ben_{m,n,k}}[A_w(H) ]  =  \left(\frac{1+z^w}{2} \right)^m {n \choose w}
\end{equation}
for $w \in [0,n]$ where $z \defeq 1- 2p$. 
The notation $[a,b]$ denotes the set of consecutive integers from $a$ to $b$.
The average weight distribution of this ensemble was 
first discussed by Litsyn and Shevelev \cite{LS02}.

If $k$ is a constant (i.e., not a function of $n$), 
this ensemble can be considered as an ensemble of sparse matrices.
In the spacial case where $k = n/2$, equal probability $1/2^{mn}$ is assigned to every matrix 
in the Bernoulli ensemble. As a simplified notation, we will denote
$
\Ren_{m,n} \defeq \Ben_{m,n,n/2},
$
where $\Ren_{m,n}$ is called the {\em random ensemble}.
Since a typical instance of $\Ren_{m,n}$ contains $\Theta(m n)$ ones, the ensemble 
can be regarded as an ensemble of dense matrices.

\subsection{Average undetected error probability of an ensemble}
For a given $m \times n$ matrix $H$, the evaluation of the undetected error probability 
$P_U(H)$ is in general computationally difficult because we need to know the weight distribution of
$C(H)$ for such evaluation. On the other hand, in some cases, we can evaluate 
the average of $P_U(H)$ for a given ensemble. Such an average probability is useful for the estimation of
the undetected error probability of a matrix which belongs to the ensemble.

Taking the ensemble average of the undetected error probability over a given ensemble $\Gen$,
we have
\begin{eqnarray} \nonumber
E_\Gen[P_U(H)] &=& E_\Gen \left[\sum_{w = 1}^n A_w(H) \epsilon^w (1-\epsilon)^{n - w} \right] \\ \label{avew}
&=& \sum_{w = 1}^n E_\Gen[A_w(H)] \epsilon^w (1-\epsilon)^{n - w} .
\end{eqnarray}
In  the above equations, $H$ can be regarded as a random variable. 
From this equation, it is evident that the average of $P_U(H)$ can be evaluated if we know the
average weight distribution of the ensemble.
For example, in the case of the random ensemble $\Ren_{m,n}$,  the average undetected error probability 
has a simple closed form.
\begin{lemma} \label{averandom}
The average undetected error probability of the random ensemble $\Ren_{m,n}$ is given by
\begin{equation}
E_{\Ren_{m,n}}[P_U(H)] =2^{-m} (1-(1-\epsilon)^n).
\end{equation}
(Proof)  By using (\ref{avew}),  we have
\begin{eqnarray} \nonumber
E_{\Ren_{m,n}}[P_U(H)] &=& \sum_{w = 1}^n E_{\Ren_{m,n}}[A_w(H)] \epsilon^w (1-\epsilon)^{n - w}  \\ \nonumber
&=& \sum_{w = 1}^n 2^{-m} {n \choose w} \epsilon^w (1-\epsilon)^{n - w}  \\
&=& 2^{-m} (1-(1-\epsilon)^n).
\end{eqnarray}
The second equality is based on the well known result \cite{Gal63}: 
\begin{equation}
E_{\Ren_{m,n}}[A_w(H)] = 2^{-m} {n \choose w}.
\end{equation}
The last equality is due to the binomial theorem. \hfill\qed
\end{lemma}

\subsection{Error exponent of undetected error probability}

For a given sequence of $(1-R)n \times n$ matrix ensembles $(n=1,2,3,\ldots,)$,
the average undetected error probability is usually an exponentially decreasing function of $n$,
where $R$ is a real number satisfying $0 < R < 1$ (called the {\em design rate}).
Thus, the exponent of the undetected error probability is of prime importance 
in understanding the asymptotic behavior of the undetected error probability.

\subsubsection{Definition of error exponent}

Let $\{ {\Gen_n} \}_{n>0} $ be a series of ensembles such that
${\Gen_n}$ consists of $(1-R)n \times n$ binary matrices.
In order to see the asymptotic behavior of the undetected error probability of
this sequence of ensembles,
it is reasonable to define the error exponent of undetected error probability in the following way:
\begin{definition}
The asymptotic error exponent of the average undetected error probability 
for a series of ensembles $\{ {\Gen_n} \}_{n>0} $ is defined by
\begin{equation}
T_{\Gen_n} \defeq \lim_{n \rightarrow \infty }\frac 1 n\log_2 E_{\Gen_n}[ P_U]
\end{equation}
if the limit exists.
\hfill\qed
\end{definition}
Henceforth we will not explicitly express the dependence of $P_U$ on $H$, writing instead  $P_U$ to denote $P_U(H)$ in all cases where     there is no fear of confusion.

The following example describes the exponent of the random ensemble.
\begin{example}
Consider the series of the random ensembles $\{\Ren_{n,(1-R)n} \}_{n>0}$.
It is easy to evaluate $T_{\Ren_{(1-R)n,n}} $:
\begin{eqnarray} \nonumber
T_{\Ren_{(1-R)n,n}}  &=&  \lim_{n \rightarrow \infty} \frac 1 n \log_2 E_{\Ren_{(1-R)n,n}}[P_U] \\ \nonumber
&=&  \lim_{n \rightarrow \infty} \frac 1 n \log_2 2^{-(1-R)n} (1-(1-\epsilon)^n) \\ \label{1mR}
&=&  -(1-R).
\end{eqnarray}
This equality implies that the average undetected error probability of the sequence of 
random ensembles behaves like
\begin{equation} \label{exponential}
E_{\Ren_{(1-R)n,n}}[P_U] \simeq 2^{-n(1-R)}
\end{equation}
if $n$ is sufficiently large. Note that 
the exponent $-(1-R)$ is independent from the crossover probability $\epsilon$.
\hfill\qed
\end{example}

\subsubsection{Error exponent and asymptotic growth rate}

The {\em asymptotic growth rate} of the average weight distribution (for simplicity henceforth 
 abbreviated as the asymptotic growth rate), which is
the basis of the derivation of the error exponent, is defined as follows.
\begin{definition}
Suppose that a series of ensembles $\{ {\Gen_n} \}_{n>0} $ is given.
If 
\[
\lim_{n \rightarrow \infty}\frac 1 n \log_2 E_{\Gen_n}[ A_{\ell n}]
\]
exists for $0 \le \ell \le 1$, then we define the {\em asymptotic growth rate} $f(\ell)$ by 
\begin{equation}
f(\ell) \defeq \lim_{n \rightarrow \infty}\frac 1 n \log_2 E_{\Gen_n}[ A_{\ell n}].
\end{equation}
The parameter $\ell$ is called  the {\em normalized weight}.
\hfill\qed
\end{definition}
From this definition, it is clear that 
\begin{equation}
 E_{\Gen_n}[ A_{\ell n}] = 2^{n(f(\ell) + o(1))},
\end{equation}
 where the notation $o(1)$ denotes terms which converge to 0 in the limit as $n$ goes to infinity.
The asymptotic growth rate of some ensembles of binary matrices
can be found in \cite{LS02}\cite{LS03}\cite{MV04}.

The next theorem gives the error exponent of the undetected error probability for 
a series of ensembles $\{ {\Gen_n} \}_{n>0}$. 
\begin{theorem}
\label{th1}
The  error exponent of $\{ {\Gen_n} \}_{n>0} $ is given by
\begin{equation}
T_{\Gen_n} 
=\sup_{0 < \ell \le 1 } [f(\ell) + \ell  \log_2\epsilon +(1 - \ell)\log_2(1 - \epsilon) ],
\end{equation}
where $f(\ell)$ is the asymptotic growth rate of $\{ {\Gen_n} \}_{n>0}$. \\
\noindent
(Proof)
Based on the definition of asymptotic growth rate, 
we can rewrite $T_{\Gen_n} $ in the form
\begin{eqnarray} \nonumber
T_{\Gen_n}
\hspace{-3mm}&=&
\hspace{-3mm}
\lim_{n \rightarrow \infty }\frac 1 n\log_2 E_{\Gen_n}[ P_U]  \\ \nonumber
&=&\hspace{-3mm}
\lim_{n \rightarrow \infty }\frac 1 n\log_2\sum_{w=1}^n E_{\Gen_n}[ A_{w}] \epsilon^{w} (1 - \epsilon)^{n - w} \\ \nonumber
&=&\hspace{-3mm}
\lim_{n \rightarrow \infty }\frac 1 n\log_2\sum_{w=1}^n 
2^{n(f(\frac{w}{n}) + K(\epsilon, n ,w)+ o(1)) } ,
\end{eqnarray}
where $K(\epsilon, n ,w)$ is defined by
\begin{equation}
K(\epsilon, n ,w) \defeq \frac{w}{n}\log_2\epsilon + \left(1 - \frac{w}{n} \right)\log_2(1 - \epsilon).
\end{equation}
Using a conventional technique for bounding summation,  
we have the following upper bound on $T_{\Gen_n}$:
\begin{eqnarray} \nonumber
T_{\Gen_n}\hspace{-3mm}
&=&\hspace{-3mm}
\lim_{n \rightarrow \infty }\frac 1 n\log_2\sum_{w=1}^n 
2^{n(f(\frac{w}{n}) + K(\epsilon, n ,w)+ o(1)) }  \\ \nonumber
&\le&\hspace{-3mm}
\lim_{n \rightarrow \infty }\frac 1 n\log_2 n \max_{w=1}^n 
2^{n(f(\frac{w}{n}) + K(\epsilon, n ,w)+ o(1)) }  \\ \nonumber
&=&\hspace{-3mm}
\lim_{n \rightarrow \infty } \max_{w=1}^n \frac 1 n\log_2
2^{n(f(\frac{w}{n}) + K(\epsilon, n ,w)+ o(1)) }  \\ \nonumber
&=&\hspace{-3mm}
\lim_{n \rightarrow \infty } \max_{w=1}^n 
\left[f\left(\frac{w}{n}\right) + K(\epsilon, n ,w) + o(1) \right] \\  \label{supeq}
&=&\hspace{-3mm}
\sup_{0 < \ell \le 1 } \left[f(\ell) + \ell  \log_2\epsilon +(1 - \ell)\log_2(1 - \epsilon) \right].
\end{eqnarray}
We can also show that $T_{\Gen_n}$ is greater than or equal to the right-hand side of
the above inequality (\ref{supeq}) in a similar manner. 
This means that the right-hand side of the inequality is asymptotically tight.

\hfill\qed
\end{theorem}

The next example discusses the case of the random ensemble.
\begin{example}
Let us again consider the series of the random ensembles given by $\{\Ren_{(1-R)n,n} \}_{n > 0}$.
These ensembles have the asymptotic growth rate $f(\ell) = h(\ell)-(1-R)$, where
the function $h(x)$ is the binary entropy function defined by
\begin{equation}
h(x) \defeq -x  \log_2 x -(1-x) \log_2 (1-x).
\end{equation}
In this case, by using Theorem \ref{th1},  we have 
\begin{equation}\label{trmn}
T_{\Ren_{(1-R)n,n}}  \hspace{-3mm}=
\sup_{0 < \ell \le 1 } [h(\ell)-(1-R) + \ell  \log_2\epsilon +(1 - \ell)\log_2(1 - \epsilon) ].
\end{equation}
Let 
\begin{equation}
D_{\ell,\epsilon} \defeq \ell \log_2 \left(\frac \ell \epsilon \right) + (1 - \ell) \log_2 \left(\frac{1-\ell}{1-\epsilon} \right).
\end{equation}
By using $D_{\ell,\epsilon} $, we can rewrite (\ref{trmn}) as
\begin{equation}
T_{\Ren_{(1-R)n,n}} =
\sup_{0 < \ell \le 1 } [-(1-R)  - D_{\ell,\epsilon} ].
\end{equation}
Since $D_{\ell,\epsilon}$ can be considered as the Kullback-Libler divergence between two probability 
distributions $(\epsilon, 1- \epsilon)$ and  $(\ell, 1- \ell)$, $D_{\ell,\epsilon}$ is always non-negative and
$D_{\ell,\epsilon} = 0$ holds if and only if $\ell = \epsilon$. Thus, we obtain
\begin{equation}
\sup_{0 < \ell \le 1 } [-(1-R)  - D_{\ell,\epsilon} ] = - (1-R),
\end{equation}
which is identical to the exponent obtained in expression (\ref{1mR}).

Let $g_\epsilon^{(rnd)}(\ell) \defeq h(\ell)-(1-R) + \ell  \log_2\epsilon +(1 - \ell)\log_2(1 - \epsilon)$.
Figure \ref{fig-random} displays the behavior of $g^{(rnd)}_{\epsilon}(\ell)$ when $R = 0.5$.
This figure confirms the result that the maximum ($\sup_{0 < \ell \le 1} g_\epsilon^{(rnd)}(\ell)= -0.5$) 
is  attained at $\ell = \epsilon$.
\begin{figure}[htbp]
  \begin{center}
  \includegraphics[scale=0.65]{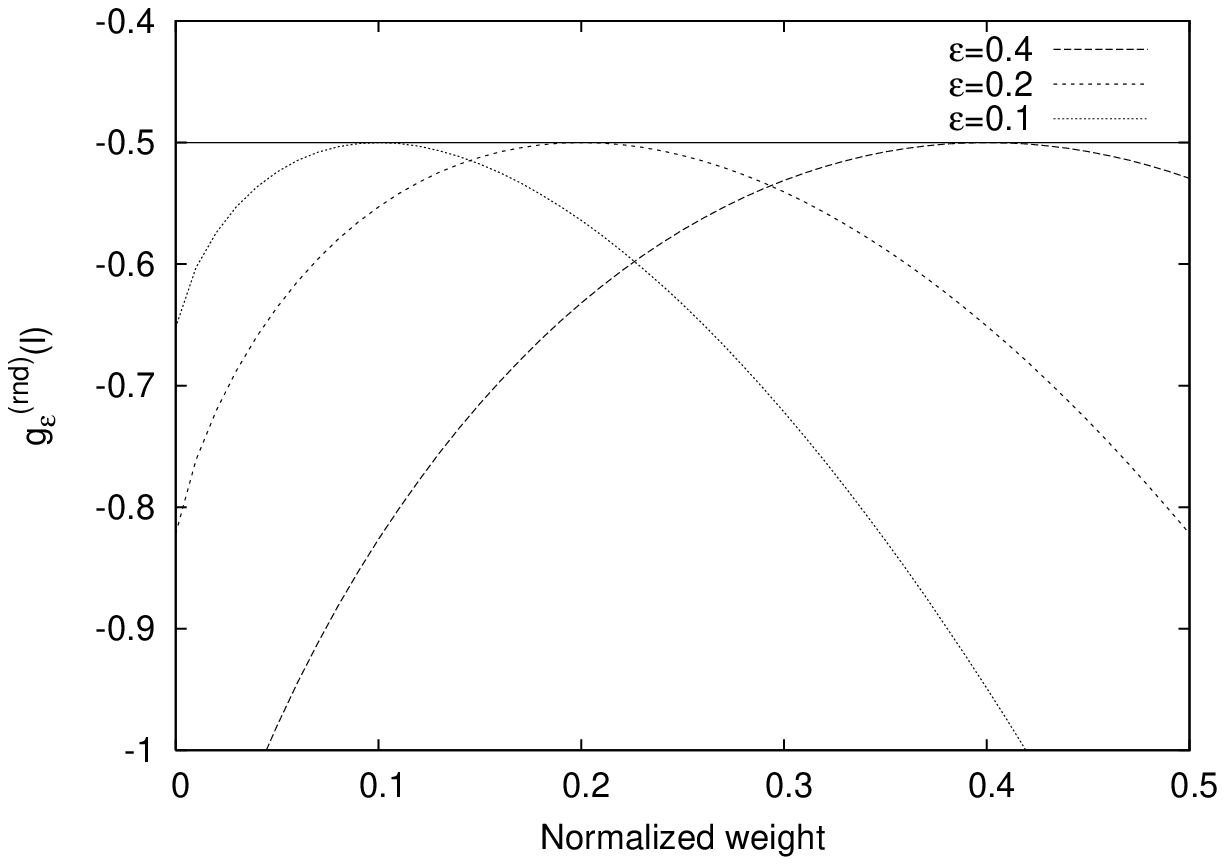} \\
\begin{flushleft}
{\tiny The curves of $g^{(rnd)}_\epsilon(\ell)$ correspond to the parameters  $\epsilon = 0.1,0.2, 0.4$ from 
left to right are presented. As a reference, line of  $-(1-R)=-0.5$ is also included in the figure.}
\end{flushleft}
      \caption{The curves of $g_\epsilon(\ell)$ for random ensembles with $R=0.5$.}
    \label{fig-random}
      \end{center}
\end{figure}
\hfill\qed
\end{example}

\subsection{Error exponent of the Bernoulli ensemble with constant $k$}

\begin{figure}[htbp]
  \begin{center}
  \includegraphics[scale=0.65]{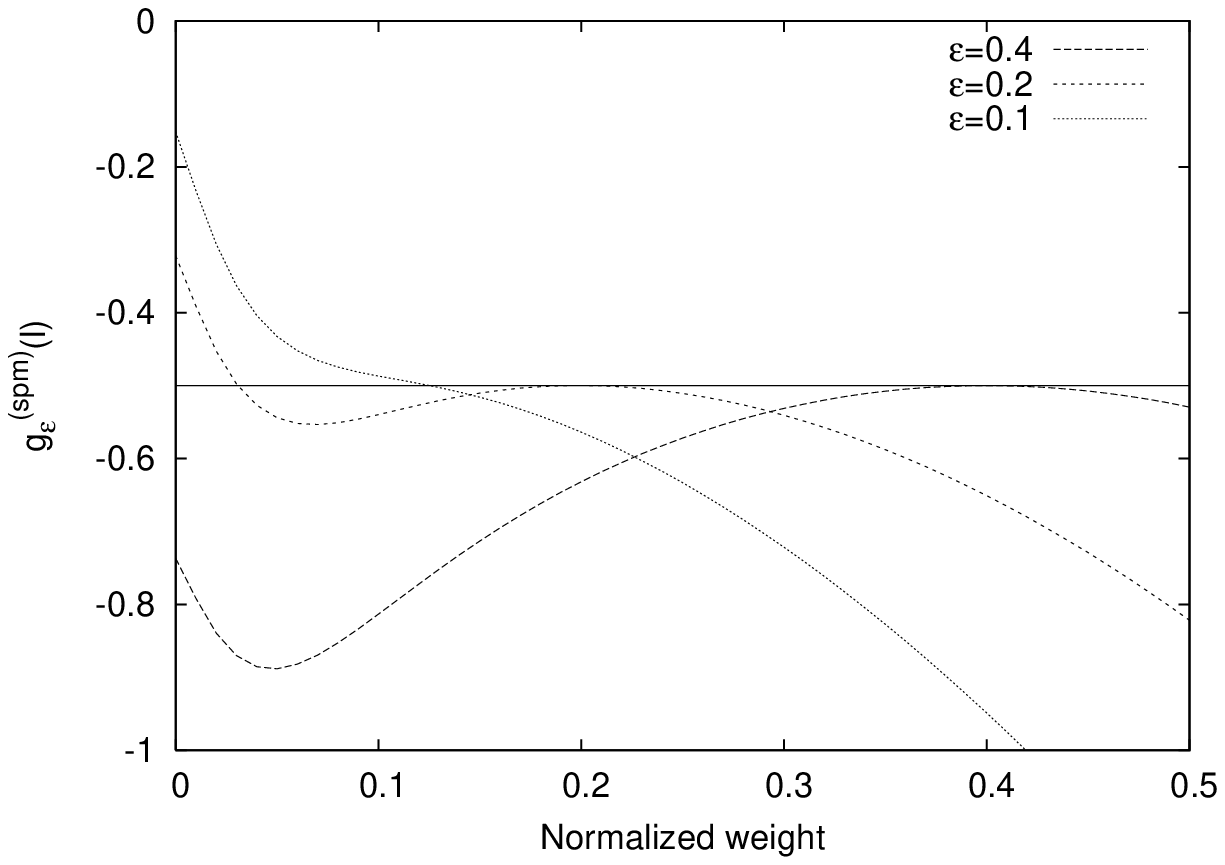} \\
{\tiny The curves of $g^{(spm)}_\epsilon(\ell)$ correspond to the parameters  $\epsilon = 0.1,0.2, 0.4$ 
are presented. The parameters $R = 0.5, k = 20$ are assumed.  As a reference, line of  $-(1-R)=-0.5$ is also included in the figure.}
      \caption{The curves of $g^{(spm)}_\epsilon(\ell)$ for Bernoulli ensembles.}
    \label{fig-sparse1}
      \end{center}
\end{figure}

The asymptotic growth rate of the Bernoulli ensemble $\Ben_{m,n,k}$ with a constant $k$ and
design rate $R$  is given by
\begin{equation}
f(\ell) = h(\ell)+ (1-R) \log_2 \left(\frac{1+e^{-2k \ell}}{2}   \right).
\end{equation}
This formula is presented in \cite{LS02}.
The error exponent of this ensemble shows a different behavior 
from that for random ensembles.
\begin{example}
Consider the Bernoulli ensemble with parameters $R = 0.5$ and $k = 20$.
Let 
\begin{eqnarray} \nonumber
g^{(spm)}_\epsilon(\ell) &\defeq& H(\ell)+ (1-R) \log_2 \left(\frac{1+e^{-2k \ell}}{2}   \right)  \\
&+& \ell  \log_2\epsilon +(1 - \ell)\log_2(1 - \epsilon).
\end{eqnarray}
\end{example}
Figure \ref{fig-sparse1} includes the curves of $g^{(spm)}_\epsilon(\ell)$ where $\epsilon = 0.1,0.2, 0.4$. In contrast to $g^{(rnd)}_\epsilon(\ell)$ of a random ensemble,  we can see that 
$g^{(spm)}_\epsilon(\ell)$ is not a concave function.  
The shape of the curve of $g^{(spm)}_\epsilon(\ell)$ depends on the crossover probability
$\epsilon$. For large $\epsilon$,  $g_\epsilon(\ell)$ takes its largest value around $\ell = \epsilon$. On the other hand,
for small $\epsilon$, $g^{(spm)}_\epsilon(\ell)$ has the supremum at $\epsilon = 0$.

\begin{figure}[htbp]
  \begin{center}
  \includegraphics[scale=0.65]{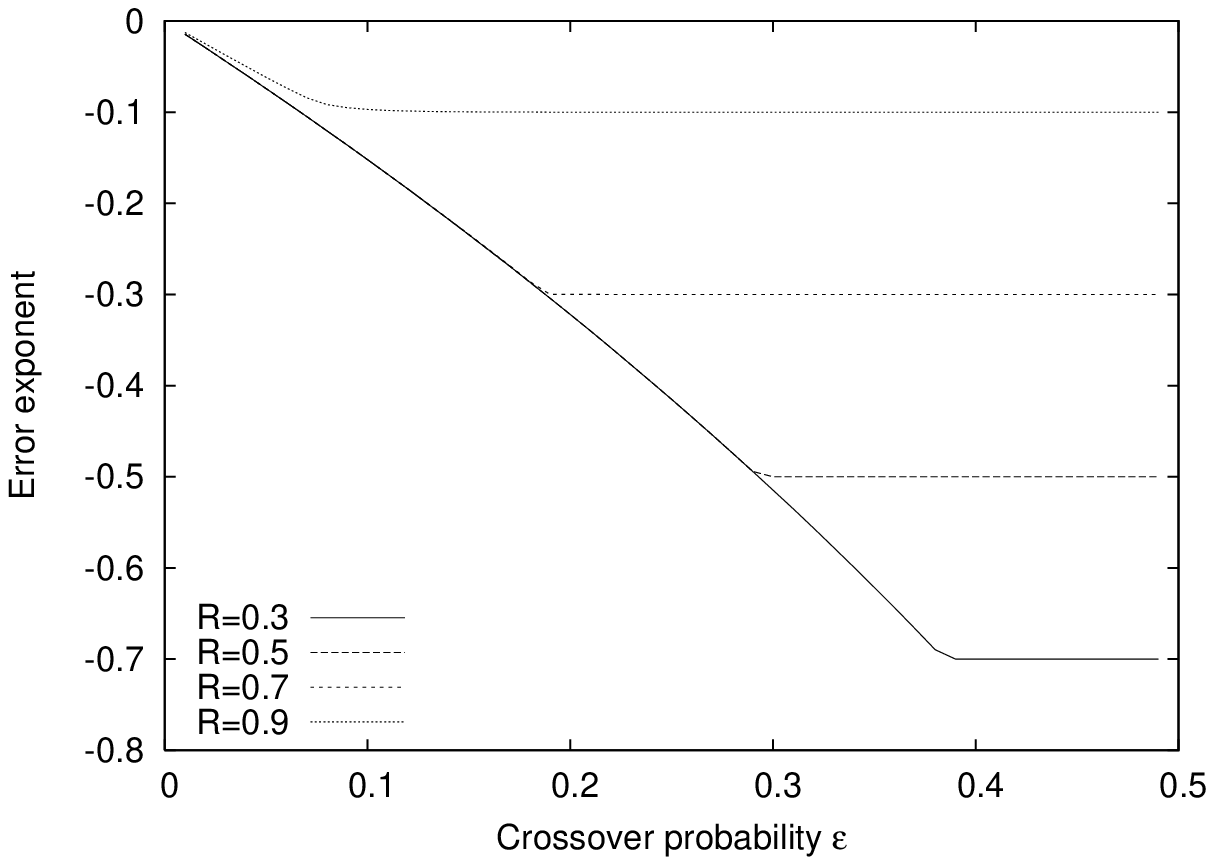} \\
{\tiny The curves of $T_{\Ben_{m,n,k}}$ correspond to the parameters  $R = 0.3,0.5, 0.7,0.9$ and $k=20$.
are presented. }
      \caption{Error exponent of Bernoulli ensemble.}
    \label{fig-sparse2}
      \end{center}
\end{figure}

Figure \ref{fig-sparse2} presents the error exponent of Bernoulli ensembles with parameters $R = 0.3, 0.5, 0.7, 0.9$ 
and $k = 20$. As an example, consider the exponent for $R=0.5$.
In the regime where $\epsilon$ is smaller than (around) 0.3, the error exponent is a monotonically decreasing function of
$\epsilon$.

The examples suggest  that a sparse ensemble has  less powerful error detection performance than 
that of a dense ensemble (such as the random ensemble) in terms of the error exponent. 
However, if the crossover probability is
sufficiently large, the difference in exponent of sparse and dense ensembles is negligible.
For example, the exponent of the Bernoulli ensemble in Fig. \ref{fig-sparse2} is almost equal to
that of the random ensemble when $\epsilon$ is larger than (around) 0.3.

\begin{figure}[htbp]
  \begin{center}
  \includegraphics[scale=0.65]{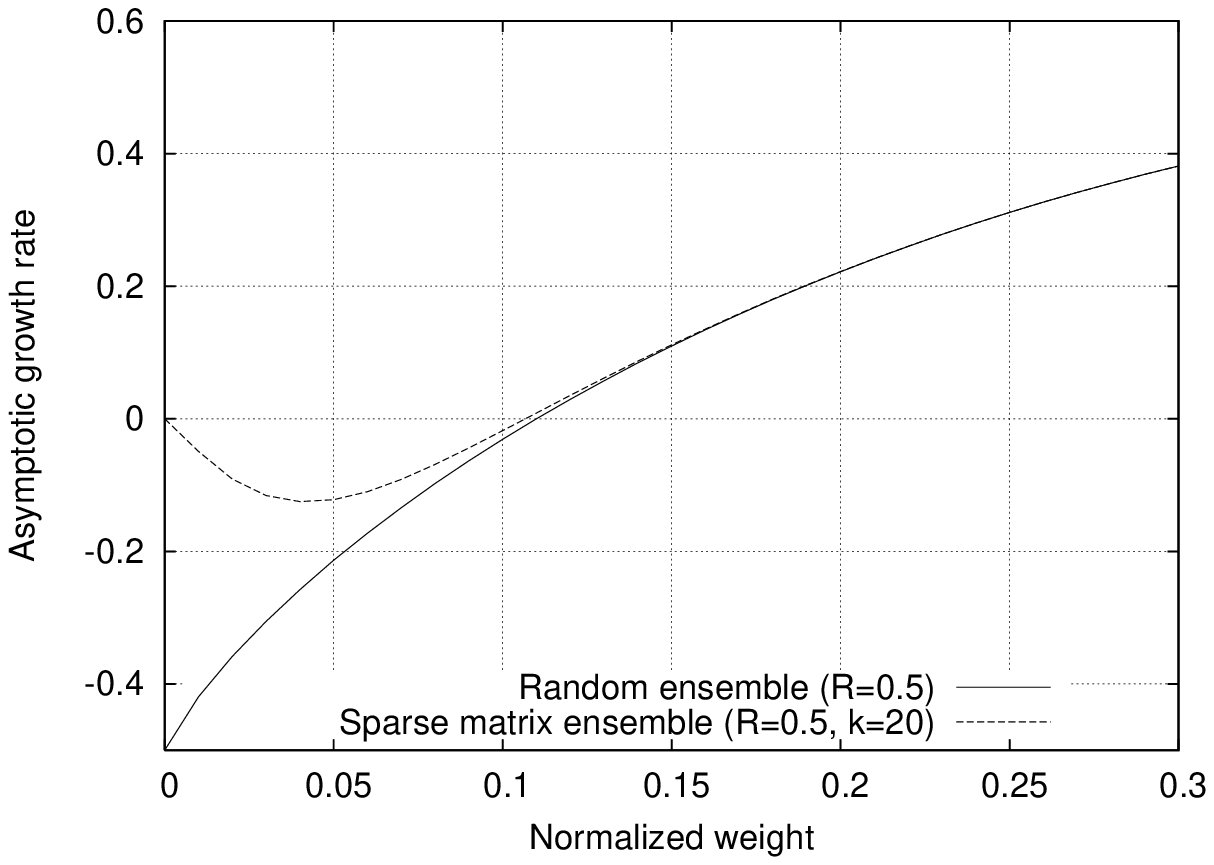} \\
      \caption{Asymptotic growth rate of a random ensemble and a Bernoulli ensemble.}
    \label{fig-agr}
      \end{center}
\end{figure}

The above properties of the error exponents of  the Bernoulli ensembles can be explained with reference to their
 average weight distributions (or asymptotic growth rate). 
Figure \ref{fig-agr} displays the asymptotic growth rates of a random ensemble and a Bernoulli ensemble.

The weight of typical error vectors is very close to  $\epsilon n$ 
when $n$ is sufficiently large.
For a large value of $\epsilon$, such as $\epsilon = 0.4$, the average weight distribution around $w = 0.4 n$, 
namely $E_\Gen[A_{0.4 n}]$,  dominates the undetected error probability. In such a range, the difference in
the average weight distributions corresponding to  the random  and the Bernoulli ensembles  is  
small. On the other hand, if the crossover probability is small, weight distributions of low weight  become
the most influential parameter. The difference in the average weight distributions of small weight 
results in a difference in the error exponent.

Note that the time complexity of the error detection operation (multiplication of received vector and a parity check matrix)
is $O(n^2)$-time for a typical instance of a random ensemble, 
and is $O(n)$-time for a typical  instance of a Bernoulli ensemble with constant $k$.
A sparse matrix offers almost  same error detection performance of a dense matrix
with linear time complexity if $\epsilon$ is sufficiently large.

\section{Variance of  undetected error probability}
In the previous section, 
we have seen that the average weight distribution plays an important role 
in the derivation of average undetected error probability.
Similarly, we need  to examine the {\em covariance of weight distribution} in order to 
analyze  the variance of undetected error probability.

\subsection{Covariance formula}

The covariance between two real-valued functions $f(\cdot), g(\cdot)$
defined on an ensemble $\G$ is given by
\begin{equation}
{\rm Cov}_{\G} [f,g] \defeq E_{\G} [ f g ] - E_{\G} [ f ] E_{\G} [ g].
\end{equation}

The next theorem forms the basis of the derivation of the variance of the undetected error probability for the Bernoulli ensemble.
The covariance of the weight distribution for the Bernoulli ensemble is given in the
following theorem.
\begin{theorem}
\label{covsparse}
The covariance of the weight distribution for the Bernoulli ensemble $\Ben_{m,n,k}$ is 
given by
\begin{eqnarray} \nonumber
&&\hspace{-15mm}{\rm Cov}_{\Ben_{m,n,k}}(A_{w_1}, A_{w_2}) \\ \nonumber
&\defeq&
\left(\frac{1+z^{w_1}}{2} \right)^m \left(\frac{1+z^{w_2}}{2} \right)^m \\ \nonumber
&\times& \hspace{-10mm}\sum_{v= \max\{0,w_1+w_2 - n\}}^{w_1}\hspace{-2mm}
 {n \choose w_1} {w_1 \choose v} {n - w_1 \choose w_2 - v} \\ \label{convformula}
&\times& \left(\left(1+ \frac{z^{w_1+w_2-2v} - z^{w_1+w_2}}{(1+z^{w_1})(1+z^{w_2})}\right)^m -1\right)
\end{eqnarray}
for $1 \le w_1 \le  w_2 \le n$ and 
\begin{equation} \label{commutative}
{\rm Cov}_{\Ben_{m,n,k}}(A_{w_1}, A_{w_2}) = {\rm Cov}_{\Ben_{m,n,k}}(A_{w_2}, A_{w_1})
\end{equation}
for $1 \le w_2 <  w_1 \le n$
where $z = 1 - 2 p$ and $p = k/n$. \\
(Proof) See Appendix. \hfill\qed
\end{theorem}

\begin{remark}
When $k = n/2$, $\Ben_{m,n,k}$ becomes the random ensemble $\Ren_{m,n}$.
We discuss this case here.

We first assume that $1 \le w_1 \le w_2\le n$.
Let  $p = 1/2$ (i.e., $k = n/2$).
In such a case, we have $z = 1 - 2p = 0$. Define $L$ by
\begin{equation}
L  \defeq \left(1+ \frac{z^{w_1+w_2-2v} - z^{w_1+w_2}}{(1+z^{w_1})(1+z^{w_2})}\right).
\end{equation}
The variable $L$ takes the following values:
\begin{equation}
L  =
\left\{
\begin{array}{ll}
1, & w_1 < w_2\\
1, & w_1 = w_2, \ v < w_1\\
2, & w_1 = w_2, \ v = w_1. 
\end{array}
\right.
\end{equation}
Substituting $z = 0$ into equation (\ref{convformula}) and using the identity (\ref{commutative}),
we get 
\begin{eqnarray} \nonumber
&&\hspace{-15mm}{\rm Cov}_{\Ren_{m,n}}(A_{w_1},A_{w_2})   \\ \label{covrandom}
&=& \left\{
\begin{array}{ll}
0, & 1 \le w_1 \ne w_2\le n\\
2^{-2m}{n \choose w} (2^m - 1), & 1 \le w_1 =w_2\le n.
\end{array}
\right.
\end{eqnarray}
Another proof of this formula is presented in \cite{acr}.
\hfill\qed
\end{remark}

\subsection{Variance of undetected error probability}

The variance of the undetected error probability is a straightforward consequence of Theorem \ref{covsparse}.
\begin{corollary}\label{spvarthreom}
The variance of the undetected error probability of the Bernoulli ensemble, $\sigma_{\Ben_{m,n,k}}^2$ is given by
\begin{eqnarray}\nonumber
\sigma^2_{\Ben_{m,n,k}} &=&  
\sum_{w_1 = 1}^n \sum_{w_2 = 1}^n {\rm Cov}_{\Ben_{m,n,k}}(A_{w_1}, A_{w_2})  \\
&\times& \epsilon^{w_1 + w_2} (1-\epsilon)^{2n - w_1 - w_2} .
\end{eqnarray}
(Proof) 
The variance of the undetected error probability $P_U$ is  given by
\begin{eqnarray} \nonumber
\sigma^2_{\Ben_{m,n,k}} &=& E_{\Ben_{m,n,k}}[(P_U - \mu)^2] \\
&=& E_{\Ben_{m,n,k}}[P_U^2] - E_{\Ben_{m,n,k}}[P_U]^2. 
\end{eqnarray}
We first consider the second moment of the undetected error probability: 
\begin{eqnarray} \nonumber
&&\hspace{-1cm} E_{\Ben_{m,n,k}}[ P_U^2] \\ \nonumber
&=& \hspace{-3mm}E_{\Ben_{m,n,k}}\left[ \left(\sum_{w = 1}^n A_w \epsilon^w (1-\epsilon)^{n - w} \right)^2 \right]
 \\ \nonumber
&=& \hspace{-3mm}E_{\Ben_{m,n,k}}\left[ \sum_{w_1 = 1}^n \sum_{w_2 = 1}^n A_{w_1} A_{w_2} \epsilon^{w_1+w_2} 
(1-\epsilon)^{2n - w_1-w_2} \right] \\
&=&  \hspace{-4mm}\sum_{w_1 = 1}^n \sum_{w_2 = 1}^n \hspace{-1mm} E_{\Ben_{m,n,k}}\left[A_{w_1} A_{w_2}\right] \epsilon^{w_1+w_2} 
(1-\epsilon)^{2n - w_1-w_2}\hspace{-1mm}.
\end{eqnarray}
The squared average undetected error probability can be expressed as
\begin{eqnarray} \nonumber
E_{\Ben_{m,n,k}}[ P_U]^2&=& \hspace{-3mm}E_{\Ben_{m,n,k}}\left[ \left(\sum_{w = 1}^n A_w \epsilon^w (1-\epsilon)^{n - w} \right) \right]^2
 \\ \nonumber
&=&  \hspace{-4mm}\sum_{w_1 = 1}^n \sum_{w_2 = 1}^n \hspace{-1mm} 
E_{\Ben_{m,n,k}}\left[A_{w_1}\right] E_{\Ben_{m,n,k}}\left[A_{w_2}  \right] \\
&\times& \epsilon^{w_1+w_2} (1-\epsilon)^{2n - w_1-w_2}\hspace{-1mm}.
\end{eqnarray}
Combining these equalities and the covariance of the weight distribution, the
variance of undetected error probability $\sigma^2_{\Ben_{m,n,k}} $ is obtained.
\hfill\qed
\end{corollary}

\begin{remark}
The covariance of the weight distribution for a given ensemble $\Ben_{m,n,k}$
is useful not only for the evaluation of the variance of $P_U$.
Let $X$ be a random variable represented by 
\begin{equation}
X = \sum_{w =0}^n \alpha(w) A_w,
\end{equation}
where $\alpha(w)$ is a real-valued function of $w$.
The covariance of the weight distribution is required more generally for the evaluation of the variance of $X$, which is given by
\begin{equation}
\sigma^2_X = \sum_{w_1=0}^n \sum_{w_2=0}^n {\rm Cov}_{\Ben_{m,n,k}}(A_{w_1}, A_{w_2}) \alpha(w_1)\alpha(w_2).
\end{equation}
A specialized version (the case where $X = P_U$) 
of this equation has been  derived in the previous corollary.
\hfill\qed
\end{remark}

\begin{example}
Let us consider the Bernoulli ensemble with $m = 1, n = 2$ and
$k = 1/2 (p = 1/4)$. 
Table \ref{r12} displays 
the weight distributions and undetected error probabilities for the 4 matrices in $\Ben_{1,2,1/2}$.
\begin{table}[htdp]
\caption{Weight distributions and undetected error probabilities}
\begin{center}
\begin{tabular}{ccccc}
\hline
\hline
$H$ &$C(H)$ & $A_1(H)$ & $A_2(H)$  & $P_U(H)$ \\
\hline
(0,0) &$\{00,01,10,11\}$ & 2 & 1 & $2\epsilon - \epsilon^2$\\
(0,1) &$\{00,10 \}$ & 1 & 0  & $\epsilon - \epsilon^2$ \\
(1,0) &$\{00,01 \}$ & 1 & 0 & $\epsilon - \epsilon^2$\\
(1,1) &$\{00,11 \}$ & 0 & 1 & $\epsilon^2$\\
\hline
\end{tabular}
\end{center}
\label{r12}
\end{table}%

From the definition of a Bernoulli ensemble, the following probability is
assigned to each matrix: 
$
P((0,0)) = 9/16, P((0,1)) = 3/16, P((1,0)) = 3/16, P((1,1)) = 1/16.
$
Combining the undetected error probabilities presented in Table \ref{r12} and the above probability assignment, 
we immediately have the first and second moments: 
\begin{eqnarray}
E_{\Ben_{1,2,1/2}}[P_U] &=& \frac 2 3 \epsilon - \frac 7 8 \epsilon^2 \\
E_{\Ben_{1,2,1/2}}[P_U^2] &=& \frac {21}{8} \epsilon^2 - \frac 3 8 \epsilon^3 + \epsilon^4.
\end{eqnarray}
From these moments, the variance can be derived:
\begin{eqnarray} \nonumber
\sigma^2_{\Ben_{1,2,1/2}}  
&=&E_{\Ben_{1,2,1/2}}[P_U^2] -E_{\Ben_{1,2,1/2}}[P_U] ^2 \\ \label{var12}
&=& \frac 3 8 \epsilon^2 - \frac 3 8 \epsilon^3 +\frac{15}{64} \epsilon^4.
\end{eqnarray}

We can also consider another route to derive the variance by using Corollary \ref{spvarthreom}.
The covariances of $\Ben_{1,2,1/2}$ are given by
\begin{eqnarray}
{\rm Cov}_{\Ben_{1,2,1/2}}(1,1) &=& 3/8 \\
{\rm Cov}_{\Ben_{1,2,1/2}}(1,2) &=& {\rm Cov}_{\Ben_{1,2,1/2}}(2,1) = 3/16 \\
{\rm Cov}_{\Ben_{1,2,1/2}}(2,2) &=& 15/64.
\end{eqnarray}
From Corollary \ref{spvarthreom},  we obtain the variance 
\begin{eqnarray} \nonumber
\sigma^2_{\Ben_{1,2,1/2}}  
&=& \sum_{w_1=1}^2 \sum_{w_2=1}^2 {\rm Cov}_{\Ben_{m,n,k}}(A_{w_1}, A_{w_2})  \\ \nonumber
&\times& \epsilon^{w_1+w_2} (1- \epsilon)^{4 - w_1-w_2} \\ \nonumber
&=& (3/8)\epsilon^2(1-\epsilon)^2 + (3/16)\epsilon^3(1-\epsilon) \\ \nonumber
&+& (3/16)\epsilon^3(1-\epsilon) + (15/64)\epsilon^4 \\ \nonumber
&=&\frac 3 8 \epsilon^2 - \frac 3 8 \epsilon^3 +\frac{15}{64} \epsilon^4,
\end{eqnarray}
that is identical to expression (\ref{var12}). \hfill\qed
\end{example}

In the case of $k = n/2$ (i.e. the case of a random ensemble), 
we can derive a closed form expression for the variance.
\begin{corollary}\label{rvartheorem}
For the random ensemble $\Ren_{m,n}$, 
the variance of the undetected error probability $P_U$ is given by
\begin{equation}
\sigma^2_{\Ren_{m,n}} =  (1-2^{-m})2^{-m}\left((\epsilon^2  + (1-\epsilon)^2)^n- (1-\epsilon)^{2n} \right).
\end{equation}
(Proof) 
The variance of undetected error probability $\sigma^2_{\Ren_{m,n}} $ can be obtained in the following way:
\begin{eqnarray} \nonumber
&& \hspace{-1cm} \sigma^2_{\Ren_{m,n}}  \\ \nonumber
&=& \hspace{-3mm}
E_{\Ren_{m,n}}[ P_U^2]-E_{\Ren_{m,n}}[ P_U]^2 \\ \nonumber
&=& \hspace{-3mm}\sum_{w_1 = 1}^n \sum_{w_2 = 1}^n {\rm Cov}_{\Ren_{m,n}}\left[A_{w_1}, A_{w_2}\right] \epsilon^{w_1+w_2} 
(1-\epsilon)^{2n - w_1-w_2}   \\ \nonumber
&=&  \hspace{-3mm}\sum_{w=1}^n(1-2^{-m})2^{-m}{n \choose w} \epsilon^{2w} (1-\epsilon)^{2n - 2w} .
\end{eqnarray}
The second equality is due to Corollary \ref{spvarthreom}.
The last equality are due to Eq. (\ref{covrandom}).
We can further simplify the expression using the binomial theorem:
\begin{eqnarray} \nonumber
\sigma^2_{\Ren_{m,n}} 
&=&  (1-2^{-m})2^{-m} \sum_{w=0}^n{n \choose w}(\epsilon^2)^{w} ((1-\epsilon)^2)^{n - w}\\ \nonumber
&-& (1-2^{-m})2^{-m}(1-\epsilon)^{2n}\\ \nonumber
&=&  (1-2^{-m})2^{-m} \\
&\times& \left((\epsilon^2  + (1-\epsilon)^2)^n- (1-\epsilon)^{2n} \right).
\end{eqnarray}
The last equality is the claim of the theorem.
\hfill\qed
\end{corollary}

The next example facilitates an understanding of how the average and the variance of $P_U$ behave.
\begin{example}
We consider the random ensemble with $m = 20, n = 40$, and 
the Bernoulli ensemble with $m = 20, n=40, k=5$ (labeled "Sparse" in Fig. \ref{fig-curve1}).
Figure \ref{fig-curve1} depicts the average undetected error probabilities of the
two ensembles. It can be observed that the average undetected error probability of the
random ensemble monotonically decreases as $\epsilon$ decreases.
In contrast, the curve for the Bernoulli ensemble has a peak around $\epsilon \simeq 0.025$.
\begin{figure}[htbp]
  \begin{center}
  \includegraphics[scale=0.65]{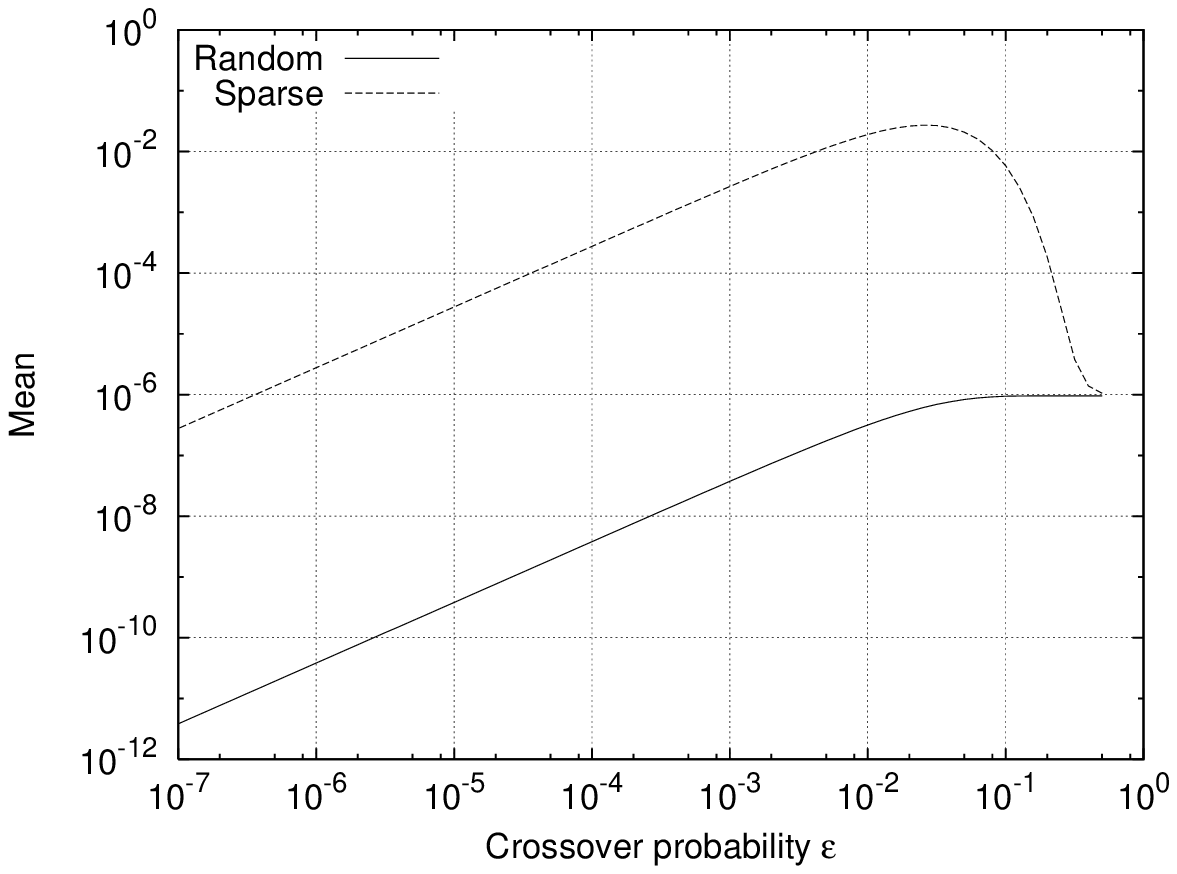} \\
{\tiny Random ensemble: $m=20, n=40$, Sparse matrix ensemble: $m=20,n=40, k=5$.}
      \caption{Average undetected error probabilities.}
    \label{fig-curve1}
      \end{center}
\end{figure}
Figure \ref{fig-curve2} shows the variance of $P_U$ for the above two ensembles. The 
two curves 
have a similar shape, but the variance of the sparse ensemble is always larger than that of the random ensemble.
\begin{figure}[htbp]
  \begin{center}
  \includegraphics[scale=0.65]{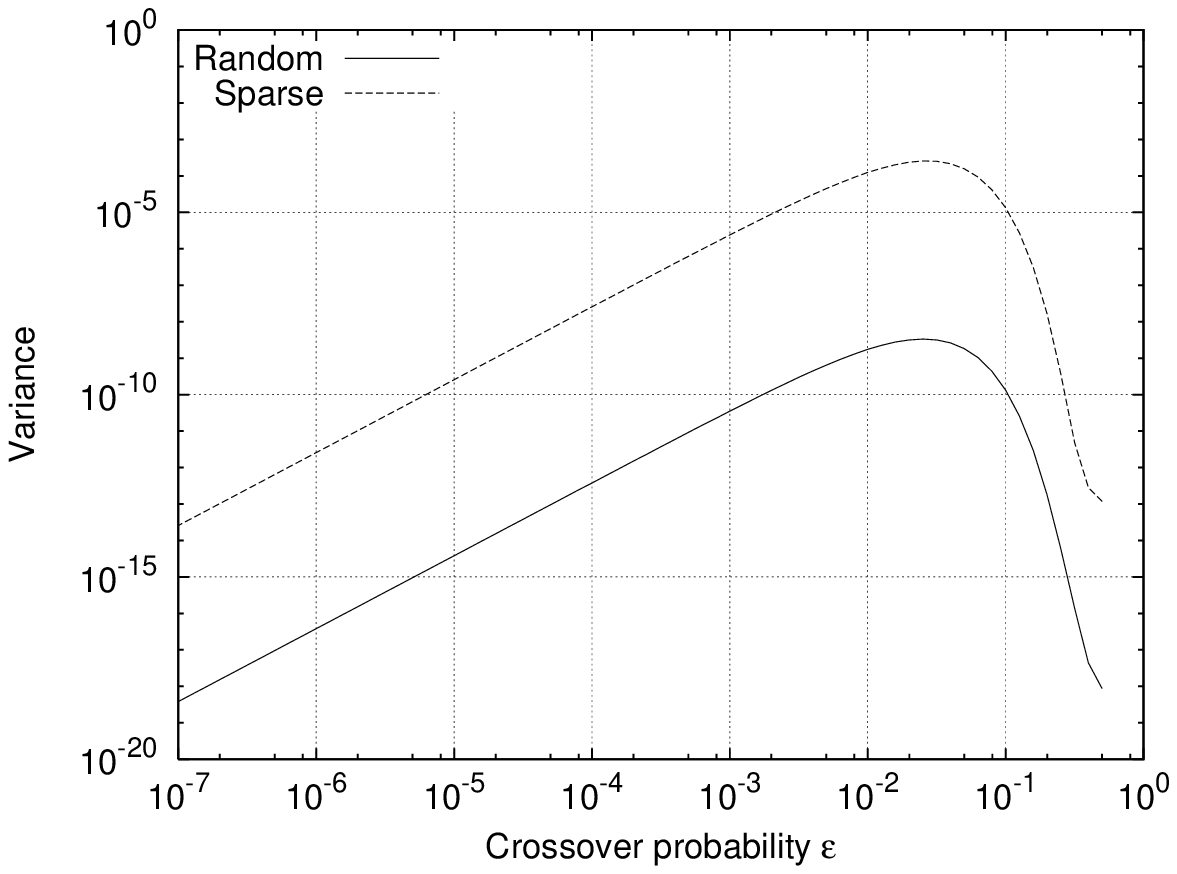} \\
{\tiny Random ensemble: $m=20, n=40$, Sparse matrix ensemble: $m=20,n=40, k=5$.}
      \caption{Variance of undetected error probability.}
    \label{fig-curve2}
      \end{center}
\end{figure}
\hfill\qed
\end{example}

\subsection{Asymptotic behavior}
We here discuss the asymptotic behavior of the covariance of the weight distribution and
the variance of $P_U$ for the Bernoulli ensemble. The following corollary explains
the asymptotic behavior of the covariance of the weight distribution.

\begin{corollary} \label{asymptcov}
Let the asymptotic growth rate of the covariance of the weigh distribution of the Bernoulli ensemble
be $T(\ell_1, \ell_2)$ defined by
\begin{equation}
T(\ell_1, \ell_2) \defeq
\lim_{n \rightarrow \infty} \frac 1 n \log_2{\rm Cov}_{\Ben_{(1-R)n ,n,k}}(A_{\ell_1 n}, A_{\ell_2 n}) 
\end{equation}
for $0 < \ell_1, \ell_2 \le 1$ and $0 < R \le 1$.
The asymptotic growth rate is given by
\begin{equation}
T(\ell_1, \ell_2) = \sup_{\max\{0, \ell_1+\ell_2 - 1\} \le  \nu \le   \ell_1}Q(\nu)
\end{equation}
for $0 < \ell_1 \le \ell_2 \le 1$ and 
\begin{equation}
T(\ell_1, \ell_2) = T(\ell_2, \ell_1)
\end{equation}
for $0 < \ell_2 < \ell_1 \le 1$
where $Q(\nu)$ is defined by
\begin{eqnarray} \nonumber
Q(\nu) &\defeq&-2(1-R) + h(\ell_1)  \\
&+& \hspace{-2mm} h\left(\frac{\nu}{\ell_1} \right)+ h\left(\frac{\ell_2 - \nu}{1-\ell_1} \right) + \hspace{-2mm}
\sup_{0 < \mu \le 1-R} \alpha(\mu,\nu).
\end{eqnarray}
The function $\alpha(\mu,\nu)$ is defined by 
\begin{eqnarray} \nonumber
&&\hspace{-10mm}\alpha(\mu,\nu)  \\ \nonumber
&\defeq& h\left(\frac{\mu}{1-R} \right)
+ \mu \log_2\left(e^{-2k(\ell_1+\ell_2-2\nu)}-e^{-2k(\ell_1+\ell_2)}\right) \\ 
&+& (1-R-\mu)\log_2\left((1+e^{-2k\ell_1})(1+e^{-2k\ell_2}) \right).
\end{eqnarray}
(Proof)
We here rewrite the covariance formula (\ref{convformula}) into asymptotic form.
By using the Binomial theorem, we have
\begin{eqnarray}\nonumber
&&\hspace{-15mm}\left(1+ \frac{z^{w_1+w_2-2v} - z^{w_1+w_2}}{(1+z^{w_1})(1+z^{w_2})}\right)^m - 1\\
&=&
\sum_{i=1}^m {m \choose i}\left(\frac{z^{w_1+w_2-2v} - z^{w_1+w_2}}{(1+z^{w_1})(1+z^{w_2})} \right)^i.
\end{eqnarray}
By using this identity, the covariance in (\ref{convformula}) can be rewritten in the following form:
\begin{eqnarray} \nonumber
&&\hspace{-6mm}{\rm Cov}_{\Ben_{m,n,k}}(A_{w_1}, A_{w_2}) \\ \nonumber
&=& 2^{-2m} \sum_{v= \max\{0,w_1+w_2 - n\}}^{w_1}\hspace{-2mm}
 {n \choose w_1} {w_1 \choose v} {n - w_1 \choose w_2 - v} \Theta,
\end{eqnarray}
where $\Theta$ is defined by
\begin{eqnarray} \nonumber
\Theta &\defeq&
\sum_{i=1}^m {m \choose i} \left(z^{w_1+w_2-2v}-z^{w_1+w_2} \right)^i \\
&\times& \left((1+z^{w_1})(1+z^{w_2}) \right)^{m-i}.
\end{eqnarray}

Letting $w_1 = \ell_1 n, w_2 = \ell_2 n, v = \nu n, m = (1-R)n$, we have
\begin{equation}
\lim_{n \rightarrow \infty}\frac{1}{n} \log_2 2^{-2m} = -2(1-R) 
\end{equation}
and
\begin{eqnarray}\nonumber
&& \hspace{-16mm }
\lim_{n \rightarrow \infty}\frac{1}{n} \log_2 {n \choose w_1} {w_1 \choose v} {n - w_1 \choose w_2 - v}  \\
&=& h(\ell_1) + h\left(\frac{\nu}{\ell_1} \right) + h\left(\frac{\ell_2 - \nu}{1-\ell_1} \right).
\end{eqnarray}
If  $k$ is a constant and $0 \le \ell \le 1$, then, making use of the identity \cite{LS02}
\begin{eqnarray} \nonumber
\lim_{n \rightarrow \infty}\left(1 - 2 \left(\frac{k}{n} \right) \right)^{\ell n} 
&=& \lim_{n \rightarrow \infty}z^{\ell n} \\
&=& e^{-2 k \ell}
\end{eqnarray}
we get 
\begin{eqnarray}
\lim_{n \rightarrow \infty}\frac{1}{n} \log_2 \Theta =  \sup_{0 < \mu \le 1-R} \alpha(\mu).
\end{eqnarray}
Combining these asymptotic expressions, the claim of the corollary is derived.
\hfill\qed
\end{corollary}

The following corollary gives the asymptotic growth rate of  the variance of 
the undetected error probability.
\begin{corollary}
The asymptotic growth rate of the variance of the undetected error is given by
\begin{equation}
\lim_{n \rightarrow \infty} \frac 1 n \log_2 \sigma^2_{\Ben_{n,(1-R)n,k}} 
=\sup_{0 < \ell_1 \le 1}\sup_{0 < \ell_2 \le 1}  S(\ell_1,\ell_2),
\end{equation}
where $S(\ell_1,\ell_2)$ is given by
\begin{eqnarray} \nonumber
S(\ell_1,\ell_2) 
&\defeq& (\ell_1 + \ell_2) \log_2 \epsilon +(2-\ell_1-\ell_2) \log_2(1-\epsilon) \\
&+& T(\ell_1, \ell_2).
\end{eqnarray}
(Proof)
It is evident that
\begin{eqnarray} \nonumber
&& \lim_{n \rightarrow \infty}\frac{1}{n} \log_2 
\left(\epsilon^{\ell_1 n + \ell_2 n} (1-\epsilon)^{2n - \ell_1 n - \ell_2 n}  \right)   \\ 
&=& (\ell_1 + \ell_2) \log_2 \epsilon +(2-\ell_1-\ell_2) \log_2(1-\epsilon).
\end{eqnarray}
holds. Combining this identity and  Corollaries \ref{spvarthreom} and \ref{asymptcov}, 
we immediately have the claim of the corollary.
\hfill\qed
\end{corollary}

\section{Appendix}
\subsubsection{Preparation of the proof}
The second moment of the weight distribution  for a given ensemble $\Gen$
is given by
\begin{eqnarray}
  \nonumber
&&\hspace{-1cm}E_{\Gen}\left[A_{w_1}A_{w_2} \right] \\ \nonumber
  &=&
  E_{\Gen}\left[
   \sum_{\ve x \in Z^{(n,w_1)}} \sum_{\ve y \in Z^{(n,w_2)}}
    I[H \ve x^t = 0^m] I[H \ve y^t = 0^m] \right].
\end{eqnarray}
for $0 < w_1, w_2 \le n$.
Since 
\[
I[H \ve x^t = 0^m] I[H \ve y^t = 0^m] = I[H \ve x^t = 0^m,H \ve y^t = 0^m], 
\]
we have
\begin{eqnarray}
  \nonumber
&&\hspace{-1cm}E_{\Gen}\left[A_{w_1}A_{w_2} \right]  \\ \nonumber
  &=&\hspace{-3mm}
  E_{\Gen}\left[
   \sum_{\ve x \in Z^{(n,w_1)}} \sum_{\ve y \in Z^{(n,w_2)}}
    I[H \ve x^t = 0^m,H \ve y^t = 0^m] \right] \\ \label{secmom}
  &=& \hspace{-7mm}
   \sum_{\ve x \in Z^{(n,w_1)}} \sum_{\ve y \in Z^{(n,w_2)}}
   E_{\Gen}\left[
    I[H \ve x^t = 0^m,H \ve y^t = 0^m] \right].
\end{eqnarray}

\begin{figure}[htbp]
  \begin{center}
  \includegraphics[scale=0.5]{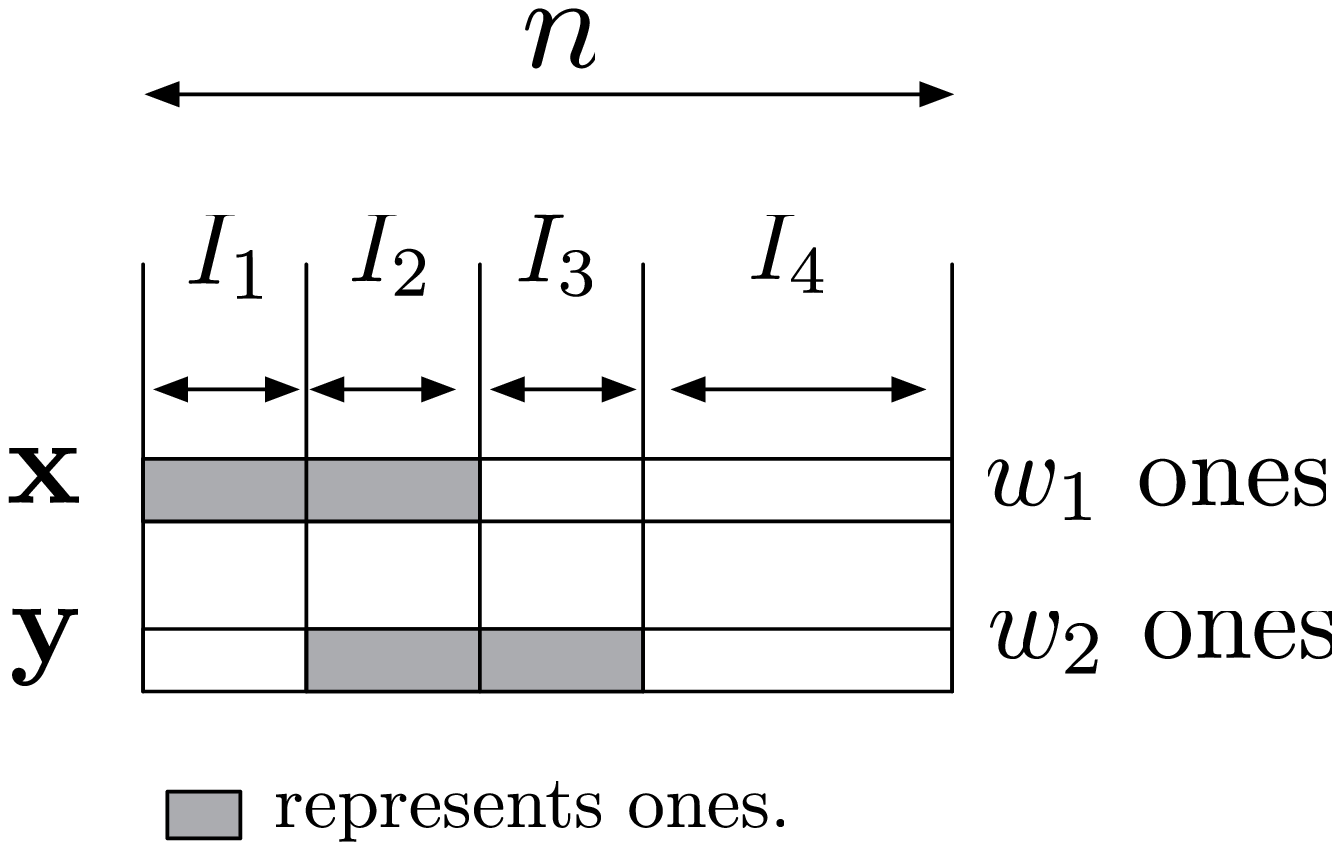} \\
      \caption{The 4 regions $I_1,I_2,I_3,I_4$.}
    \label{fig-regions}
      \end{center}
\end{figure}

We  here encounter a problem of evaluating probability of occurrence of 
both $H \ve x^t = 0^m$ and  $H \ve y^t = 0^m$.
In preparation to solve this problem, we will introduce some notation:
\begin{definition}
For a given pair $(\ve x, \ve y) \in Z^{(n,w_1)} \times Z^{(n,w_2)}$,
the index sets $I_1,I_2,I_3,I_4$ are defined as follows:
\begin{eqnarray}
I_1 &\defeq& \{k \in [1,n]: x_k = 1, y_k = 0 \} \\
I_2 &\defeq& \{k \in [1,n]: x_k = 1, y_k = 1 \} \\
I_3 &\defeq& \{k \in [1,n]: x_k = 0, y_k = 1 \} \\
I_4 &\defeq& \{k \in [1,n]: x_k = 0, y_k = 0 \},
\end{eqnarray}
where $\ve x=(x_1,x_2,\ldots,x_n)$ and $\ve y=(y_1,y_2,\ldots,y_n).$
These regions are illustrated in Fig.\ref{fig-regions}.
The size of each index set is denoted by  $i _k = \# I_k (k = 1,2,3,4)$.
Let $\ve h=(h_1,h_2,\ldots,h_n)$ be a binary $n$-tuple. 
The partial weight of $\ve h$ corresponding to an index set $I_k(k=1,2,3,4)$
is denoted by $w_k(\ve h)$, namely 
\begin{equation}
  w_k(\ve h) = \# \{j \in I_k: h_j = 1\}.
\end{equation}
\hfill\qed 
\end{definition}

Since the index sets are mutually exclusive, the equation $i_1+i_2+i_3+i_4 = n$ holds 
and $i_2$ can take an integer value in the following range:
\begin{equation}
\max\{w_1+w_2-n,0  \} \le i_2\le \min\{w_1,w_2 \}.
\end{equation}
The size of each index set can be expressed as 
$i_1 = w_1 - i_2$, $i_3 = w_2 - i_2$, 
$i_4= n - (w_1+w_2 - i_2)$.

\subsection{Proof of Lemma \ref{covsparse} (Covariance of the Bernoulli ensemble)}

Let  $\ve x \in Z^{(n,w_1)}$ and $\ve y \in Z^{(n,w_2)}$ be binary vectors
satisfying  $w_1 \le w_2$.
In this proof, we first prove the following equality:
\begin{eqnarray} \nonumber
&&\hspace{-10mm} E_{{\Ben}_{n,m,k}} [I[H \ve x^t = 0, H \ve y^t = 0] ]    \\ \label{xyprob}
&=&\left(\frac{1+z^{w_1} + z^{w_2} + z^{w_1 + w_2  - 2 v} }{4} \right)^m
\end{eqnarray}
where
$v = \#(\supp(\ve x) \cap \supp(\ve x))$, $z = 1 - 2 p$ and $p = k/n$. 
The support set $\supp(\ve v)$ is defined by
\begin{equation}
\supp(\ve v) \defeq \{i \in [1,n]: v_i \ne 0 \},
\end{equation}
where $\ve v = (v_1,v_2,\ldots,v_n)$.

We need to consider the following three cases:
Case (i): $0 < i_2 < w_1$ (i.e., the intersection of $\supp(\ve x)$ and $\supp(\ve y)$ is not empty 
but $\supp(\ve y)$ does not include $\supp(\ve x)$),
Case (ii): $i_2 = 0$ (i.e.,  the intersection of $\supp(\ve x)$ and  $\supp(\ve y)$ is empty),
Case (iii): $i_2 = w_1$ (i.e.,  $\supp(\ve y)$ includes $\supp(\ve x)$).

We first study Case (i).
Suppose that a binary $n$-tuple $\ve h$ is generated from a Bernoulli source with $Pr[h_i = 1] = p (i \in [1,n])$.
Recall that $p$ is defined by $p = k/n$.
In this case, $\ve h \ve x^t = 0, \ve h \ve y^t = 0$ holds if and only if
$w_{i}(\ve h)\  \mbox{is even}$ for $i=1,2,3$ or  $w_{i}(\ve h)\  \mbox{is odd}$ for $i=1,2,3$.

It is well known that a binary vector $(t_1,t_2,\ldots, t_u)$ generated from a Bernoulli source 
has even weight with probability $(1+ (1-2 q)^u)/2$, where $q$ is the probability   that
$t_i (i \in [1,u])$ takes 1 \cite{Gal63}.  The probability  that $(t_1,t_2,\ldots, t_u)$ has an odd weight is 
given by $(1- (1-2 q)^u)/2$.
For example, the probability   that $w_{1}(\ve h)$ becomes even is $(1+z^{w_1})/2$
where $z = 1  - 2p$. 

Based on the above argument, we can write
the probability $Pr[\ve h \ve x^t = 0, \ve h \ve y^t = 0]$  as a function of $z$:
\begin{eqnarray}  \nonumber
&&\hspace{-10mm} Pr[\ve h \ve x^t = 0, \ve h \ve y^t = 0] \\ \nonumber
\hspace{-3mm}
&=&\hspace{-3mm} \frac{(1+z^{i_1})(1+z^{i_2})(1+z^{i_3})+(1 -z^{i_1})(1-z^{i_2})(1-z^{i_3}) }{8} \\
&=&\hspace{-3mm}\frac{1+z^{w_1} + z^{w_2} + z^{w_1 + w_2  - 2 v} }{4}.
\end{eqnarray}
where $v \defeq i_2$.

We next consider Case (ii). For this case, $v=i_2$ is assumed to be zero.
In this case, $\ve h \ve x^t = 0, \ve h \ve y^t = 0$ holds if and only if both $w_1(\ve h)$ and $w_3(\ve h)$ are even.
The probability   that $\ve h$ satisfies $\ve h \ve x^t = 0$ and $ \ve h \ve y^t = 0$ under the condition $i_2 = 0$ 
is given by
\begin{eqnarray} \nonumber
&&\hspace{-28mm} Pr[\ve h \ve x^t = 0, \ve h \ve y^t = 0] \\ \nonumber
&=&\left(\frac{1+z^{i_1}}{2}\right) \left(\frac{1 +z^{i_3}}{2} \right) \\ \nonumber
&=&\left(\frac{1+z^{w_1}}{2}\right) \left(\frac{1 +z^{w_2}}{2} \right) \\
&=&\frac{1+z^{w_1} + z^{w_2} + z^{w_1 + w_2  - 2 v} }{4}.
\end{eqnarray}

Finally we consider Case (iii).
Assume the case   $v = i_2 = w_1, \ve x \ne \ve y$.
In this case, $\ve h \ve x^t = 0, \ve h \ve y^t = 0$ holds if and only if both $w_2(\ve h)$ and $w_3(\ve h)$ are even.
The probability $Pr[\ve h \ve x^t = 0, \ve h \ve y^t = 0]$
under the condition $v = w_1, \ve x \ne \ve y$ is thus given by
\begin{eqnarray} \nonumber
&&\hspace{-22mm}Pr[\ve h \ve x^t = 0, \ve h \ve y^t = 0]  \\ \nonumber
&=& \left(\frac{1+z^{i_2}}{2}\right) \left(\frac{1 +z^{i_3}}{2} \right) \\ \nonumber
&=& \frac{1+z^{w_1}+z^{w_2}  + z^{w_2 - w_1}   }{4} \\
&=& \frac{1+z^{w_1}+z^{w_2}  + z^{w_1 + w_2 - 2 v}   }{4}. 
\end{eqnarray}
We next consider the case  $\ve x = \ve y$.
For this case, we also have 
\begin{eqnarray} \nonumber
&&\hspace{-22mm}Pr[\ve h \ve x^t = 0, \ve h \ve y^t = 0]  \\ \nonumber
&=& \frac{1+x^{w_1}}{2} \\
&=&\frac{1+z^{w_1} + z^{w_2} + z^{w_1 + w_2  - 2 v} }{4}.
\end{eqnarray}

In summary, for any cases (Cases (i), (ii), (iii)), 
\begin{equation}
Pr[\ve h \ve x^t = 0, \ve h \ve y^t = 0]  = \frac{1+z^{w_1} + z^{w_2} + z^{w_1 + w_2  - 2 v} }{4}
\end{equation}
holds. Since the rows of parity check matrices in ${\Ben}_{n,m,k}$ can be independently chosen, we obtain 
Eq. (\ref{xyprob}) in the following way:
\begin{eqnarray} \nonumber
&& \hspace{-10mm} E_{{\Ben}_{n,m,k}} [I[H \ve x^t = 0, H \ve y^t = 0] ] \\ \nonumber
&=& Pr[H \ve x^t = 0, H \ve y^t = 0 ]  \\ \nonumber
&=&Pr[\ve h \ve x^t = 0, \ve h \ve y^t = 0]^m \\ 
&=&\left(\frac{1+z^{w_1} + z^{w_2} + z^{w_1 + w_2  - 2 v} }{4} \right)^m.
\end{eqnarray}

Combining (\ref{secmom}) and (\ref{xyprob}), we have
\begin{eqnarray}
  \nonumber
&&\hspace{-1cm}E_{{\Ben}_{n,m,k}}\left[A_{w_1}A_{w_2} \right]  \\  \nonumber
 &=& \hspace{-7mm}
   \sum_{\ve x \in Z^{(n,w_1)}} \sum_{\ve y \in Z^{(n,w_2)}}
   E_{{\Ben}_{n,m,k}}\left[
    I[H \ve x^t = 0^m,H \ve y^t = 0^m] \right] \\ \nonumber
 &=& \hspace{-7mm}
   \sum_{\ve x \in Z^{(n,w_1)}} \sum_{\ve y \in Z^{(n,w_2)}}
\left(\frac{1+z^{w_1} + z^{w_2} + z^{w_1 + w_2  - 2 v} }{4} \right)^m \\ \nonumber
&=& \hspace{-2mm}\sum_{v= \max\{0,w_1+w_2 - n\}}^{w_1} {n \choose w_1} {w_1 \choose v} {n - w_1 \choose w_2 - v} \\ \label{2nd}
&\times& \left(\frac{1+z^{w_1} + z^{w_2} + z^{w_1 + w_2  - 2 v} }{4} \right)^m.
\end{eqnarray}
Since
\begin{equation}
E_{{\Ben}_{n,m,k}}\left[A_{w} \right]  = {n \choose w} \left(\frac{1+z^w}{2} \right)^m
\end{equation}
holds \cite{LS02}, we thus have
\begin{eqnarray} \nonumber
&&\hspace{-12mm} E_{{\Ben}_{n,m,k}}\left[A_{w_1} \right] E_{{\Ben}_{n,m,k}}\left[A_{w_2} \right] \\  \nonumber
&=& {n \choose w_1}{n \choose w_2} \left(\frac{1+z^{w_1}}{2} \right)^m \left(\frac{1+z^{w_2}}{2} \right)^m \\ \nonumber
&= &\sum_{v= \max\{0,w_1+w_2 - n\}}^{w_1} {n \choose w_1} {w_1 \choose v} {n - w_1 \choose w_2 - v} \\  \label{exp2}
&\times& \left(\frac{1+z^{w_1} + z^{w_2} + z^{w_1 + w_2 } }{4} \right)^m. 
\end{eqnarray}
The last equality is due to the following combinatorial identity:
\begin{equation}
\sum_{v= \max\{0,w_1+w_2 - n\}}^{w_1} {n \choose w_1} {w_1 \choose v} {n - w_1 \choose w_2 - v}
= {n \choose w_1} {n \choose w_2}.
\end{equation}
We are ready to derive the covariance of weight distributions for the case $w_1 \le w_2$.
Substituting (\ref{2nd}) and (\ref{exp2}) into 
\begin{eqnarray} \nonumber
&&\hspace{-10mm}{\rm Cov}_{\Ben_{m,n,k}}(A_{w_1}, A_{w_2}) \\ \nonumber
&=&E_{{\Ben}_{n,m,k}}\left[A_{w_1}A_{w_2} \right] - 
E_{{\Ben}_{n,m,k}}\left[A_{w_1} \right] E_{{\Ben}_{n,m,k}}\left[A_{w_2} \right],
\end{eqnarray}
we have (\ref{convformula}) in the claim part of the Theorem.
Since the definition of covariance is commutative, 
${\rm Cov}_{\Ben_{m,n,k}}(A_{w_1}, A_{w_2}) = {\rm Cov}_{\Ben_{m,n,k}}(A_{w_2}, A_{w_1})$ holds
if $w_1 > w_2$.
\hfill\qed

\section*{Acknowledgment}
This work was partly supported by the Ministry of Education, Science, Sports
and Culture, Japan, Grant-in-Aid for Scientific Research on Priority Areas
(Deepening and Expansion of Statistical Informatics) 180790091.

\end{document}